\documentclass[conference]{IEEEtran}
\IEEEoverridecommandlockouts
\usepackage{cite}
\usepackage{amsmath,amssymb,amsfonts}
\usepackage{textcomp}
\usepackage{xcolor}
\def\BibTeX{{\rm B\kern-.05em{\sc i\kern-.025em b}\kern-.08em
    T\kern-.1667em\lower.7ex\hbox{E}\kern-.125emX}}

\usepackage{epsfig}
\usepackage{color}
\usepackage{amsmath} % real numbers, natural numbers, ...
\usepackage{amssymb}
\usepackage{amsxtra}
\usepackage{enumitem}

\usepackage[T1]{fontenc}% optional T1 font encoding
\usepackage{setspace}

\usepackage{amsmath}
\usepackage[cmintegrals]{newtxmath}
\usepackage{graphicx}
\graphicspath{{Figures/}}
\usepackage{url}
\usepackage{cite}
\usepackage{algorithm}
\usepackage{algpseudocode}

\makeatletter
\def\BState{\State\hskip-\ALG@thistlm}
\makeatother

\usepackage{epstopdf}
\usepackage{amsmath}
\usepackage{amsxtra}
\usepackage{amstext}
\usepackage{amssymb}
\usepackage{latexsym}
\usepackage{dsfont} % for \mathds{N}
\usepackage{color}
\usepackage{multirow}
\usepackage{float}
\usepackage{hyperref}

\usepackage{tabularx,colortbl}

\usepackage{lscape}

\usepackage{algorithm}
\usepackage{algpseudocode}

\usepackage{units}
\usepackage{cases}
\usepackage[ subrefformat=parens,labelformat=parens, caption=false, font=footnotesize]{subfig}
\usepackage{mathtools}
\mathtoolsset{showonlyrefs} %only reference numbered equations

\newcommand{\mbf}[1]{\mathbf{#1}}
\newcommand{\nth}[1]{{#1}{\text{th}}}

\newcommand{\abs}[1]{\left|{#1}\right|}
\newcommand{\norm}[1]{\left\|{#1}\right\|}
\DeclareMathOperator*{\argmax}{arg\,max}   % In your preamble
\DeclareMathOperator*{\argmin}{arg\,min}

\usepackage{mathtools}

\newcommand{\ML}{\mathrm{ML}}

\newcommand{\ZF}{\mathrm{ZF}}

\newcommand{\MMSE}{\mathrm{MMSE}}
\newcommand{\GRAND}{\mathrm{GRAND}}

\newcommand{\Prob}{\mathrm{Pr}}

\newfont{\bb}{msbm10 scaled 1100}
\newcommand{\CC}{\mbox{\bb C}}

\newcommand{\RR}{\mbox{\bb R}}

\newcommand{\ZZ}{\mbox{\bb Z}}
\newcommand{\FF}{\mbox{\bb F}}

\begin{document}

\title{GRAND for Fading Channels using \\Pseudo-soft Information}

\author{\IEEEauthorblockN{Hadi~Sarieddeen}
\IEEEauthorblockA{\textit{Research Laboratory of Electronics} \\
\textit{Massachusetts Institute of Technology}\\
Cambridge, MA 02139, USA \\
hadisari@mit.edu}
\and
\IEEEauthorblockN{Muriel M{\'e}dard}
\IEEEauthorblockA{\textit{Research Laboratory of Electronics} \\
\textit{Massachusetts Institute of Technology}\\
Cambridge, MA 02139, USA \\
medard@mit.edu}
\and
\IEEEauthorblockN{Ken. R. Duffy}
\IEEEauthorblockA{\textit{Hamilton Institute} \\
\textit{Maynooth University}\\
Ireland \\
ken.duffy@mu.ie}
\thanks{The project or effort depicted was or is sponsored by the Defense Advanced Research Projects Agency under Grant number HR00112120008. The content of the information does not necessarily reflect the position or policy of the Government, and no official endorsement should be inferred.

To appear in the IEEE GLOBECOM 2022 proceedings.}
}

%\author{Hadi~Sarieddeen,
%        Muriel M{\'e}dard,~\IEEEmembership{Fellow,~IEEE,}
%       and~Ken.~R.~Duffy% <-this % stops a space
%\thanks{H. Sarieddeen and M. Medard are with the Research Laboratory of Electronics, Massachusetts Institute of Technology, Cambridge, MA 02139, USA. 
%(e-mail: hadisari@mit.edu; medard@mit.edu). Ken. R. Duffy is with the Hamilton Institute, Maynooth University, Ireland. (e-mail: ken.duffy@mu.ie). 
%}% <-this % stops a space
%\thanks{Manuscript received xxx, xxx; revised xxx, xxx.}
%}

\maketitle

\begin{abstract}

Guessing random additive noise decoding (GRAND) is a universal maximum-likelihood decoder that recovers code-words by guessing rank-ordered putative noise sequences and inverting their effect until one or more valid code-words are obtained. This work explores how GRAND can leverage additive-noise statistics and channel-state information in fading channels. Instead of computing per-bit reliability information in detectors and passing this information to the decoder, we propose leveraging the colored noise statistics following channel equalization as pseudo-soft information for sorting noise sequences. We investigate the efficacy of pseudo-soft information extracted from linear zero-forcing and minimum mean square error equalization when fed to a hardware-friendly soft-GRAND (ORBGRAND). We demonstrate that the proposed pseudo-soft GRAND schemes approximate the performance of state-of-the-art decoders of CA-Polar and BCH codes that avail of complete soft information. Compared to hard-GRAND, pseudo-soft ORBGRAND introduces up to $\unit[10]{dB}$ SNR gains for a target $10^{-3}$ block-error rate.

\end{abstract}


\section{Introduction}
\label{sec:introduction}

Guessing random additive noise decoding (GRAND) \cite{Duffy8437648,Duffy8630851} has recently been proposed as a capacity-achieving universal maximum-likelihood (ML) channel-code decoder. Instead of identifying the transmitted code-word, GRAND identifies the most likely noise sequence that corrupts the code-word. In particular, GRAND rank-orders putative noise sequences and successively reverses the received signal's noise effects from most likely to least likely to recover candidate transmitted words. As the noise sequences are ordered in decreasing likelihood, leveraging information on noise and channel models, it provably follows that the first code-word to be recovered is the ML decoded solution, even for channels with memory and with the absence of interleaving. Furthermore, the guesswork \cite{Arikan481781,Beirami8522043} in GRAND is computationally feasible for all moderate redundancy codes because, in most cases, the Shannon entropy rate of noise is less than that of information symbols \cite{Duffy8630851}.

The NP-completeness of ML decoding \cite{Berlekamp1055873} has traditionally compelled practical code-specific decoding paradigms. For example, soft-detection cyclic redundancy check (CRC)-assisted successive cancellation list (CA-SCL) decoding exploits the structure of CRC-assisted polar (CA-Polar) codes for computational efficiency \cite{Tal7055304}. However, supporting the requirements of diverse emerging wireless communications applications \cite{rajatheva2020white}, such as Internet of things (IoT) and virtual reality, requires a plethora of enablers that range from ultra-reliable low-latency communication (URLLC) \cite{Durisi7529226} to high-frequency ultra-broadband connectivity \cite{Sarieddeen9514889}. A paradigm shift to practical universal decoders that are efficient for different code lengths and rates is thus desirable in efforts to realize these technologies. Early universal soft-detection near-ML decoders for linear codes adopted a list-decoding principle \cite{Gazelle567691,Valembois1291728}, in which the conditional likelihood of the received signal is computed to only a restricted list of code-words. GRAND is a novel practical universal decoder suitable for block-code constructions of moderate redundancy--even unstructured code-books. GRAND's modularity and computational efficiency translate into significant hardware footprint reduction \cite{Riaz9567867,Abbas9195254}.

Leveraging soft symbol reliability information enhances decoding accuracy \cite{Kaneko605601,Fossorier412683}. Soft information for GRAND can be a one-bit mask that designates whether a channel is reliable \cite{Duffy8849297}. Fully utilizing real-valued channel outputs as soft information is also possible in soft-GRAND (SGRAND) \cite{Solomon9149208}, where noise-sequence likelihoods are inferred from the reliability of demodulated symbols. Ordered reliability bits GRAND (ORBGRAND) \cite{Duffy9414615} achieves the decoding accuracy of SGRAND in a parallelizable and hardware-friendly algorithm, using code-book-independent quantization of soft information. However, generating soft information in detectors incurs a computational overhead, and passing such information with high resolution to a decoder on every channel use is a baseband processing bottleneck. A recent proposal to incorporate information from channel state information (CSI) into GRAND constructs a mask that designates bits as reliable or not depending on whether a fading coefficient exceeds a threshold to be optimized \cite{abbas2022grand} (an extension to \cite{Duffy8849297}). Fully integrating fading information into GRAND is not yet studied.

This work proposes pseudo-soft GRAND schemes that do not avail of soft-detection bit-reliability information, leveraging CSI and colored additive noise (CAN) statistics alongside hard demapped symbols instead. In particular, noise coloring after equalization is dictated by channel fading coefficients that can be retained over many frames in wideband systems with low computational overhead. We consider the zero-forcing (ZF) and minimum mean square error (MMSE) channel equalizers and study the suitability of their resulting pseudo-soft information for GRAND. By integrating knowledge of CAN, the proposed GRAND schemes favor keeping noise bursts and foregoing interleaves \cite{Duffy8630851,An9322303} or whitening filters.

The paper is organized as follows. The problem formulation is first presented in Sec.~\ref{sec:sysmodel}. Then, the proposed pseudo-soft GRAND for fading channels is detailed and analyzed in Sec.~\ref{sec:proposed}. Performance evaluation is reported in Sec.~\ref{sec:results}. Regarding notation, bold upper case, bold lower case, and lower case letters correspond to matrices, vectors, and scalars, respectively. Scalar and vector $\text{L}_2$ norms are denoted by $\abs{\cdot}$ and $\norm{\cdot}$. $\mathsf{E}[\cdot]$, $(\cdot)^{T}$, and $(\cdot)^{*}$, stand for the expected value, transpose, and conjugate transpose. $\mbf{I}_M$ is an identity matrix of size $M$, $\FF_u$ denotes a Galois field with $u$ elements, $\Prob(\cdot)$ is the probability function, and $\odot$ is the Hadamard product.

\section{System Model and Problem Formulation}
\label{sec:sysmodel}

\subsection{System Model}
\label{sec:system}

We consider a communication system of equivalent baseband input-output relation, $\mbf{y} \!=\! \mbf{h}\!\odot\!\mbf{x} \!+\! {\mbf{n}}$,
where $\mbf{y}\!=\![{y}_{1}\!\cdots\! {y}_{i}\!\cdots\! {y}_{M}^{}]^{T}\!\in\!\CC^{M\times1}$ is the received symbol vector, $\mbf{h}\!=\![{h}_{1}\!\cdots\! {h}_{i}\!\cdots\! {h}_{M}^{}]^{T}\!\in\!\CC^{M\times1}$ is the vector of channel coefficients, $\mbf{x}\!=\![x_{1}\!\cdots\! x_{i}\!\cdots\! x_{M}^{}]^{T}\!\in\!\CC^{M\times1}$ is a frame of transmitted symbols, and ${\mbf{n}}\!=\![{n}_{1}\!\cdots\! {n}_{i}\!\cdots\! {n}_{M}^{}]^{T}\!\in\!\CC^{M\times1}$ is the additive complex-Gaussian noise vector of zero mean and per-symbol variances $\left(\mathsf{E}[{n}_i {n}_i^{*}]\!=\!\sigma_i^{2}\right)$. Furthermore,  each information symbol, $x_{i}$, belongs to a normalized complex constellation, $\mathcal{X}_{i}$ $\left(\mathsf{E}[x_{i}^{*}x_{i}^{}]\!=\!1\right)$. We shall consider quadrature amplitude modulation (QAM) of different orders in this paper. For a single-input single-output (point-to-point) communication link, $y_i = h_i x_i + n_i$, $i\!\in\! \{1,\cdots,M\}$, is an independent channel use, where $h_i$ can represent Rayleigh (when the $h_i$s are complex Gaussian with zero mean) or Rician (when the $h_i$s are complex Gaussian with possibly non-zero mean) fading, for example.

The bit-representation of $x_i$ is $\mbf{c}_{i}\!=\![c_{i,1}\cdots c_{i,j}\cdots c_{i,q_i}]^{T}\!\in\! \FF_2^{q_i}$, where $q_i\!=\!\lceil{\log_2(\abs{\mathcal{X}_i})\rceil}$. For simplicity, we consider a uniform modulation type over symbols; $q_i\!=\!q$, $\forall i$. The bit-representation of $\mbf{x}$ is thus $\mbf{c}\!=\![\mbf{c}_{1}\!\cdots\! \mbf{c}_{i} \!\cdots\! \mbf{c}_{M}]^{T}\!\in\!\FF_2^N$, where $N\!=\! \sum_{i=1}^{M} q_i\!=\!Mq$. We assume $\mbf{c}$ to be a code-word encoded with an error correcting code $\alpha\!:\! \FF_2^K \!\rightarrow\! \FF_2^{N}$, of code-rate $R\!=\!K/N$. A code-book $\mathcal{C}\!\triangleq\!\{ \mbf{c}\!:\! \mbf{c} \!=\! \alpha(\mbf{b}), \mbf{b}\!\in\!\FF_2^K\}$ includes all possible code-words, where $\mbf{b}$ is the string of bits corresponding to the  information payload we seek to transmit.  We denote by $\mbf{v}\!=\! [\sigma_{1,1}^2 \!\cdots\!\sigma_{i,j}^2 \!\cdots\! \sigma_{M,q}^2]\!\in\!{\RR^+}^N\!$ a vector of second-order noise statistics per bit. Assuming perfect knowledge of the $h_i$s at the receiver, a hard detector, $\bar{\alpha}\!:\!\CC^M\!\rightarrow\!\bar{\mathcal{X}}$, equalizes the channel and recovers a symbol vector, $\hat{\mbf{x}}$, from $\mbf{y}$, where $\bar{\mathcal{X}}\!=\!\mathcal{X}_i^M$ is the finite $M$-dimensional lattice of all possible modulated symbols. A demapper recovers a word, $\hat{\mbf{c}}$, from $\hat{\mbf{x}}$.

\subsection{Problem Formulation}
\label{sec:problem}

\begin{algorithm}[b]
\caption{Soft GRAND}\label{alg:SGRAND}
\begin{algorithmic}[1]
\Require Soft-inf. $\mbf{\Lambda}$; demapped bits $\hat{\mbf{c}}$; noise matrix $\mbf{W}$; ordered noise-generating function $\Pi$; abandonment threshold $B$
\Ensure Decoded $\hat{\mbf{c}}^{\GRAND}$
\State $\mbf{t} \gets \text{sort}\left(\mbf{W}\times\abs{\mbf{\Lambda}}\right)$ \Comment{sort in increasing order}
\State $k \gets 0$;
\While{$k < 2^N$}
\State $k \gets k+1$; $\mbf{w} \gets \Pi(\mbf{t}(k))$ \Comment{$\nth{k}$ likely noise sequence}
\If{$\hat{\mbf{c}}\ominus \mbf{w} \in \mathcal{C}$ \textbf{or} $k=B$}
    \State $\hat{\mbf{c}}^{\GRAND} \gets \hat{\mbf{c}}\ominus \mbf{w}$
    %\State $\hat{\mbf{b}}^{\GRAND} \gets \alpha^{-1}(\hat{\mbf{c}}^{\GRAND})$  \Comment{information bits}
    \State \textbf{return} $\hat{\mbf{c}}^{\GRAND}$
\EndIf
\EndWhile
\end{algorithmic}
\end{algorithm}

In the traditional view, an ML decoder computes the conditional probability of the demapped word, $\hat{\mbf{c}}$, for each of the $2^K$ code-words, $\mbf{c}$, in $\mathcal{C}$. The $\mbf{c}$ with the highest conditional likelihood of transmission given what was received is the ML solution, $\mbf{c}^{\ML} \!=\! \argmax \{ \Prob \left( \hat{\mbf{c}}\!\mid\! \mbf{c}\right)\!:\! \mbf{c} \!\in\! \mathcal{C}\}$.
%\begin{equation}\label{eq:ML_dec}
%\mbf{c}^{\ML} \in \argmax \{ \Prob \left( \hat{\mbf{c}}\mid \mbf{c}\right): \mbf{c} \in \mathcal{C}\}.
%\end{equation}
Instead of searching through $\mathcal{C}$ for code-words, GRAND searches putative, not necessarily memoryless, noise effect sequences that corrupt $\mbf{c}$, $\mbf{w}\!=\![w_{1,1}\!\cdots\! w_{i,j}\!\cdots\! w_{M,q}^{}]^{T}\!\in\!\FF_2^N$, with non-increasing probability. We express the interaction between a bit and the channel's  noise effect on that bit through the function $\oplus$; $\hat{\mbf{c}} = \mbf{c}\oplus \mbf{w}$. We can write $\Prob \left(\hat{\mbf{c}}\!\mid\! \mbf{c}\right) \!=\! \Prob \left(\hat{\mbf{c}} \!=\!  \mbf{c}\!\oplus\! \mbf{w}\right)$. Therefore,
\begin{equation}\label{eq:GRAND}
\hat{\mbf{c}}^{\GRAND} = \argmax \{ \Prob \left(\mbf{w}= \hat{\mbf{c}} \ominus \mbf{c}\right): \mbf{c} \in \mathcal{C}\},
\end{equation}
where $\ominus$ is the inversion of the noise effect. The receiver creates, usually in an online and highly parallelizable fashion, a list of noise sequences sorted in decreasing order of likelihood, and queries until the first code-word hit (block-code syndrome computations), $\mbf{w}= \hat{\mbf{c}} \ominus \mbf{c}$. GRAND is thus a ML decoder that returns $\hat{\mbf{c}}^{\GRAND}$; information bits are retrieved as $\hat{\mbf{b}}^{\GRAND} \!=\! \alpha^{-1}(\hat{\mbf{c}}^{\GRAND})$. Since the entropy of noise is typically much smaller than that of information bits, GRAND is of low complexity. GRAND efficiency is further guaranteed by abandoning guesswork after a fixed computational cut-off \cite{Duffy8630851}, a simplification that can maintain optimal error exponents.   
%the search for candidate noise sequences

Soft-GRAND algorithms accept, in addition to $\hat{\mbf{c}}$, a vector of bit-reliability information, $\mbf{\Lambda}\!=\![\lambda_{1,1}\cdots \lambda_{i,j}\cdots \lambda_{M,q}^{}]^{T}\!\in\!\RR^N$. We can generate a weight metric by multiplying $\mbf{w}$ by $\abs{\mbf{\Lambda}}$; noise sequences with smaller weights are more likely to occur.
Hence, in soft-GRAND (Algorithm \ref{alg:SGRAND}), $\mbf{\Lambda}$ rank-orders candidate noise sequences. Let $\mbf{W}\!\in\!\FF_2^{2^N\times N}$ be a noise matrix containing in its rows all possible noise sequences, and let $\Pi: \ZZ^+ \rightarrow \FF_2^N$ be a noise-retrieving function over $\mbf{W}$. Then, $\mbf{t}=\text{sort}\left(\mbf{W}\times\abs{\mbf{\Lambda}}\right)\!\in\!{\ZZ^+}^{2^N}$ is a vector of sorted (increasing order) noise-sequence indices, and $\mbf{w} = \Pi(\mbf{t}(k))$ retrieves the $\nth{k}$ likely noise sequence. However, populating noise sequences in a single matrix is not hardware-friendly nor computationally efficient. Alternatively, SGRAND \cite{Solomon9149208} achieves the benchmark optimal decoding performance via a dynamic algorithm that recursively constructs a max-heap for each combination of reliabilities in $\mbf{\Lambda}$ to generate $\mbf{w}$ vectors with increasing likelihoods. In a highly computationally lightweight alternative, ORBGRAND \cite{Duffy9414615} builds a bit permutation map based on the decreasing rank order of bit reliability to generate a pre-determined series of putative noise queries.
However, $\mbf{\Lambda}$ is not always available, as soft-output detectors are computationally demanding. Communicating $\mbf{\Lambda}$ between the detector and decoder also consumes considerable bandwidth. We aim at generating pseudo-soft reliability information using only CSI and arbitrary CAN statistics.

\section{Proposed Soft GRAND for Fading Channels}
\label{sec:proposed}

\subsection{Soft-detection information in fading channels}
\label{sec:soft_inf}

We first detail the derivation of soft information under different data detectors. Reexpressing the system model as
\begin{equation}\label{eq:sys2}
\mbf{y} = \mbf{H}\mbf{x} + {\mbf{n}},
\end{equation}
$\mbf{H}\!=\!\text{diag}(\mbf{h})\!\in\!\CC^{M\times M}$ has diagonal elements, $\mbf{H}(i,i)\!=\!h_i$, and zero off-diagonal elements. Following ML detection, soft information in the form of log-likelihood ratios (LLRs) is accumulated in a vector $\mbf{\Lambda}^{\ML}\!=\![\lambda_{1,1}^{\ML}\cdots \lambda_{i,j}^{\ML}\cdots \lambda_{Mq}^{\ML}]^{T}\!\in\!\RR^{N\times1}$. The ML detector \cite{jalden2004maximum} aims at maximizing the probability of correctly estimating $\mbf{x}$ by a candidate $\hat{\mbf{x}}$,
\begin{equation}\label{eq:prob}
    \Prob\left(\mbf{x} = \hat{\mbf{x}} | \mbf{y},\mbf{H}\right) = \frac{\Prob\left(\mbf{x} = \hat{\mbf{x}}\right) f_{\mbf{y}|\mbf{x},\mbf{H}}\left(\mbf{y}|\mbf{x} = \hat{\mbf{x}},\mbf{H}\right) }{f_{\mbf{y}|\mbf{H}}\left(\mbf{y}|\mbf{H}\right)},
\end{equation}
which is maximized by the $\hat{\mbf{x}}$ that maximizes $f_{\mbf{y}|\mbf{x},\mbf{H}}\left(\mbf{y}|\mbf{x} = \hat{\mbf{x}},\mbf{H}\right)$, where $f_{\mbf{y}|\mbf{x},\mbf{H}}\left(\mbf{y}|\mbf{x} = \hat{\mbf{x}},\mbf{H}\right)$ and $f_{\mbf{y}|\mbf{H}}\left(\mbf{y}|\mbf{H}\right)$ are probability density functions of $\mbf{y}$ given $\left(\mbf{x},\mbf{H}\right)$ and $\mbf{H}$, respectively. From \eqref{eq:sys2}, 
\begin{equation}\label{eq:mlsys}
f_{\mbf{y}|\mbf{x},\mbf{H}}\left(\mbf{y}|\mbf{x} = \hat{\mbf{x}},\mbf{H}\right) = f_{\mbf{n}}\left(\mbf{y} - \mbf{H}\hat{\mbf{x}}\right).
\end{equation}
Furthermore, for Gaussian noise, 
\begin{equation}\label{eq:Gauss}
f_{\mbf{n}}\left(\mbf{n}\right) = \frac{1}{\pi^N\det\left(\Gamma\right)}e^{-\left(\mbf{y} - \mbf{H}\hat{\mbf{x}}\right)^*\Gamma^{-1}\left(\mbf{y} - \mbf{H}\hat{\mbf{x}}\right)},
\end{equation}
where $\Gamma \!=\! \text{diag}\left(\mbf{v}\right)$. In the particular case of $\sigma_{i,j}\!=\!\sigma, \forall i,j$,
\begin{equation}\label{eq:Gauss2}
f_{\mbf{n}}\left(\mbf{n}\right) = \frac{1}{\left(\pi\sigma^2\right)^N}e^{-\frac{1}{\sigma^2}\norm{\mbf{n}}^2} = \frac{1}{\left(\pi\sigma^2\right)^N}e^{-\frac{1}{\sigma^2}\norm{\mbf{y} - \mbf{H}\hat{\mbf{x}}}^2}.
\end{equation}
Therefore, a ML detector exhaustively searches the lattice $\bar{\mathcal{X}}$, computing $\abs{\mathcal{X}_1}\times\abs{\mathcal{X}_i}\times\cdots\times\abs{\mathcal{X}_M}$ Euclidean distance metrics, to solve for $\hat{\mbf{x}}^{\ML} \!=\! \argmin_{\mbf{x} \in \bar{\mathcal{X}}}\norm{\mbf{y} - \mbf{H}\mbf{x}}^{2}$. Furthermore, the ML LLR of the $\nth{j}$ bit of the $\nth{i}$ symbol is computed as
\begin{equation}\label{eq:LLR_ML1}
    \lambda_{i,j}^{\ML} \!=\! \log \frac{\Prob\left( c_{i,j} \!=\! 1, \mbf{y},\mbf{H}\right)}{\Prob\left( c_{i,j} \!=\! 0, \mbf{y},\mbf{H}\right)} \!=\! \log \frac{\sum_{\mbf{x} \in \bar{\mathcal{X}}^{i,j,1}}e^{-\frac{1}{\sigma^2}\norm{\mbf{y} - \mbf{H}\mbf{x}}^2}}{\sum_{\mbf{x} \in \bar{\mathcal{X}}^{i,j,0}}e^{-\frac{1}{\sigma^2}\norm{\mbf{y} - \mbf{H}\mbf{x}}^2}},
\end{equation}
assuming uniform priors, where $\bar{\mathcal{X}}^{i,j,1}\!\triangleq\!\{\mbf{x} \!\in\! \bar{\mathcal{X}}: c_{i,j}\!=\!1\}$ and $\bar{\mathcal{X}}^{i,j,0}\!\triangleq\!\{\mbf{x} \!\in\! \bar{\mathcal{X}}: c_{i,j}\!=\!0\}$ are subsets of symbol vectors in $\bar{\mathcal{X}}$, having in the corresponding $\nth{j}$ bit of the $\nth{i}$ symbol a value of $1$ and $0$, respectively. Using the Jacobian-logarithm approximation, $\log\sum_{r}{\exp(a_{r})}\approx\max_{r}\{a_{r}\}$, optimal LLRs in the log-max sense \cite{Ivanov7436797} are expressed as \cite{8186206Sarieddeen}
\begin{equation}\label{eq:LLR_ML2}
  \lambda_{i,j}^{\ML} \approx \frac{1}{\sigma^{2}} \left(\min_{\mbf{x} \in \bar{\mathcal{X}}^{i,j,1}}{\norm{\mbf{y} - \mbf{H}\mbf{x}}^{2}} - \min_{\mbf{x} \in \bar{\mathcal{X}}^{i,j,0}}{\norm{\mbf{y} - \mbf{H}\mbf{x}}^{2}} \right).
\end{equation}
Furthermore, in point-to-point links, \ref{eq:LLR_ML2} can be executed at a much-reduced complexity over decoupled $x_i$ symbols,
\begin{equation}\label{eq:LLR_ML3}
  \lambda_{i,j}^{\ML} = \frac{1}{\sigma_i^{2}} \left(\min_{x_i \in \mathcal{X}_i^{j,1}}{\abs{y_i - h_i x_i}^{2}} - \min_{x_i \in \mathcal{X}_i^{j,0}}{\abs{y_i - h_i x_i}^{2}} \right),
\end{equation}
where $\mathcal{X}_i^{j,1}\!\triangleq\!\{x_i \!\in\! \mathcal{X}_i: c_{i,j}\!=\!1\}$ and $\mathcal{X}_i^{j,0}\!\triangleq\!\{x_i \!\in\! \mathcal{X}_i: c_{i,j}\!=\!0\}$ are subsets of symbols in the one-dimensional constellation, $\mathcal{X}_i$, having a $\nth{j}$ bit of $1$ and $0$, respectively. 

Soft information can also be extracted from linear detectors \cite{studer2011asic}, which are near-optimal in point-to-point systems. In particular, a ZF detector equalizes the channel by multiplying by its pseudo-inverse, resulting in $\mbf{\hat{y}}^{\ZF} \!=\! \left(\mbf{H}^{*}\mbf{H}\right)^{-1} \mbf{H}^{*}\mbf{y}$, which reduces to $\hat{y}_i^{\ZF} \!=\! h_i^{-1}y_i \!=\! x_i + h_i^{-1}n_i$ for a diagonal $\mbf{H}$. Alternatively, a linear MMSE detector accounts for the signal-to-noise ratio (SNR), $\mathsf{SNR}\!=\!1/\sigma^{2}$ for a uniform $\sigma$, and computes an equalized output, $\mbf{\hat{y}}^{\MMSE} \!=\! \left(\mbf{H}^{*}\mbf{H}\!+\!\left(1/\mathsf{SNR}\right)\mbf{I}_{2}\right)^{-1} \mbf{H}^{*}\mbf{y}$, which reduces to $\hat{y}_i^{\MMSE} \!=\! \left(h_i^{*} h_i\!+\!\sigma_i^{2}\right)^{-1}h_i^*y_i$ at the symbol-level of a point-to-point system. Therefore, the ZF and MMSE LLRs in $\mbf{\Lambda}^{\ZF}$ and $\mbf{\Lambda}^{\MMSE}$ are calculated per symbol as
%over decoupled search spaces \textcolor{red}{MM: This may not be so clear, we have not mentioned decoupling of what and what the search space is}, where
\begin{equation}\label{eq:LLR_ZF}
\lambda_{i,j}^{\ZF} = \frac{1}{{\sigma_{i}^{\ZF}}^2} \left(\min_{x_i\in \mathcal{X}_i^{j,1}} \abs{\hat{y}_i^{\ZF} - x_i}^{2} - \min_{x_i\in \mathcal{X}_i^{j,0}} \abs{\hat{y}_i^{\ZF} - x_i}^{2}\right),
\end{equation}
\begin{equation}\label{eq:LLR_MMSE}
\lambda_{i,j}^{\MMSE} \!=\! \frac{1}{{\sigma_{i}^{\MMSE}}^2}\! \left(\min_{x_i\in \mathcal{X}_i^{j,1}} \abs{\hat{y}_i^{\MMSE} \!-\! x_i}^{2} \!-\! \min_{x_i\in \mathcal{X}_i^{j,0}} \abs{\hat{y}_i^{\MMSE} \!-\! x_i}^{2}\right),
\end{equation}
where ${\sigma_{i}^{\ZF}}^2\!=\!\sigma_i^{2} \!\left(h_i^{*} h_i\right)^{-1}$ and ${\sigma_{i}^{\MMSE}}^2\!=\!\sigma_i^{2} \!\left(h_i^{*} h_i\!+\!\sigma_i^{2}\right)^{-1}$.% are the scaled noise variances. 
%For an additive white Gaussian noise (AWGN) system, $\mbf{y}=\mbf{x}+\mbf{n}$ ($h_i \!=\! 1$, $\forall i$), the ML, ZF, and MMSE detectors are identical. (wrong)

%Furthermore, the hard-output detected vector can be inferred from the sign of the soft output; with negative bipolar mapping, $\hat{c}_{i,j}^{\ML/\ZF/\MMSE} =\left( \text{sign}\left(\lambda_{i,j}^{\ML/\ZF/\MMSE}\right)+1\right)/2$.%a 0 is mapped to positive LLR, and a 1 is mapped to negative LLR. Therefore:
%\begin{equation}\label{eq:demap}
%    \hat{c}_{i,j}^{\ML/\ZF/\MMSE} =\left( \text{sign}\left(\lambda_{i,j}^{\ML/\ZF/\MMSE}\right)+1\right)/2.
%\end{equation}
%\textcolor{red}{MM: Is the rest of this paragraph truly necessary?}Furthermore, in the particular case of a complex AWGN system with binary phase-shift keying (BPSK), where $x_i\!\in\!\mathcal{X}=(1/\sqrt{2})\{+1+1i,-1-1i\}$ and $n_i \!\sim\! \mathcal{CN}(0,\sigma_i^{2})$, soft information is computed per bit as 
 %    \begin{equation}\label{eq:LLR_Slice}
 %       \lambda_{i,j} = \frac{1}{\sigma_i^{2}}\left(\abs{{y}_i - (+1+1i)}^{2} - \abs{{y}_i - (-1-1i)}^{2}\right).
 %    \end{equation}

\begin{figure*}[!ht]
    \centering
    \subfloat[Rank-ordered soft information at 10dB; BPSK modulation.]{\label{fig:llrrician} \includegraphics[width=0.48\linewidth]{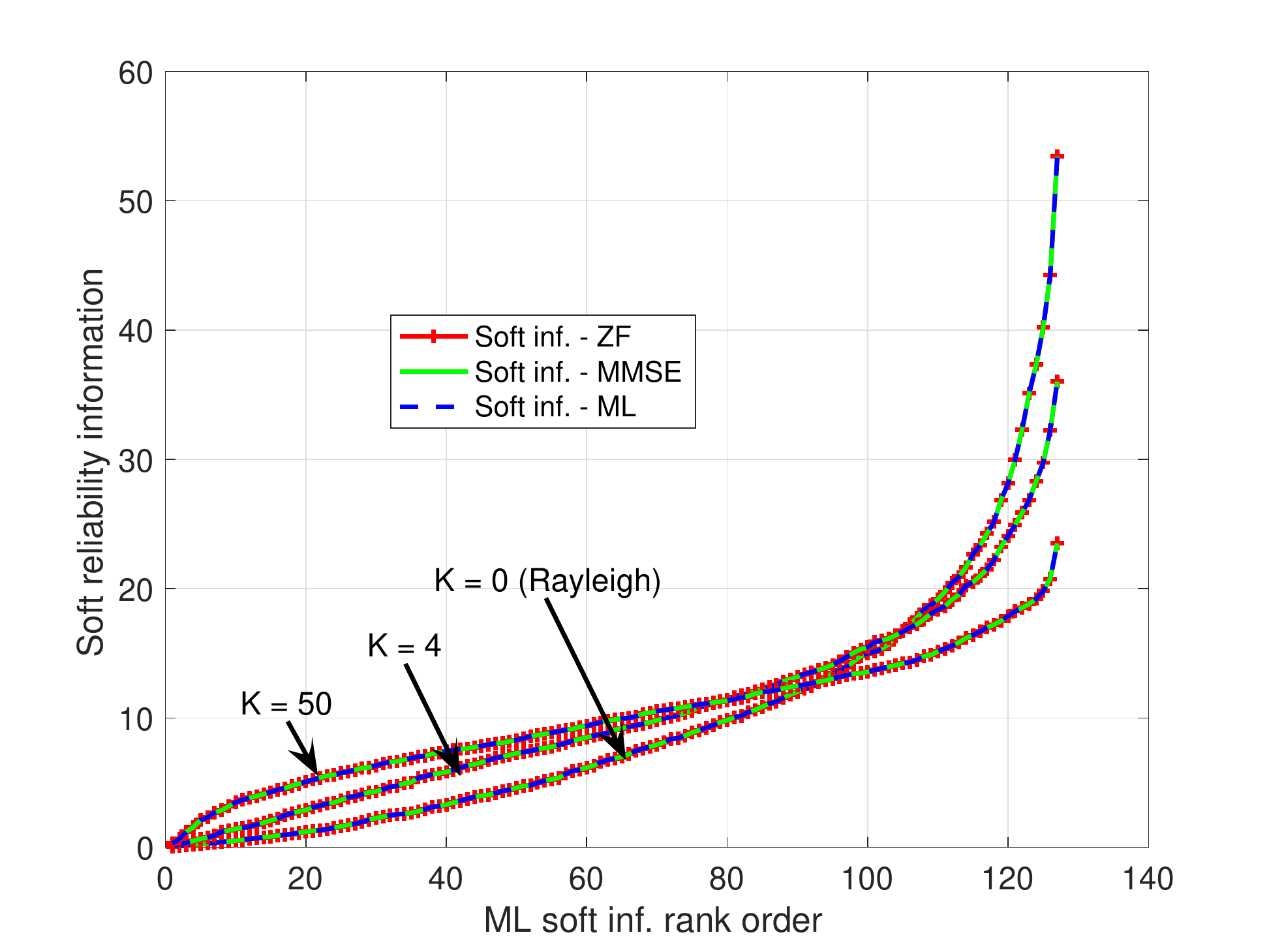}} 
    \hfill
    \subfloat[Rank-ordered pseudo-soft information at 10dB.]{\label{fig:noiserician}\includegraphics[width=0.48\linewidth]{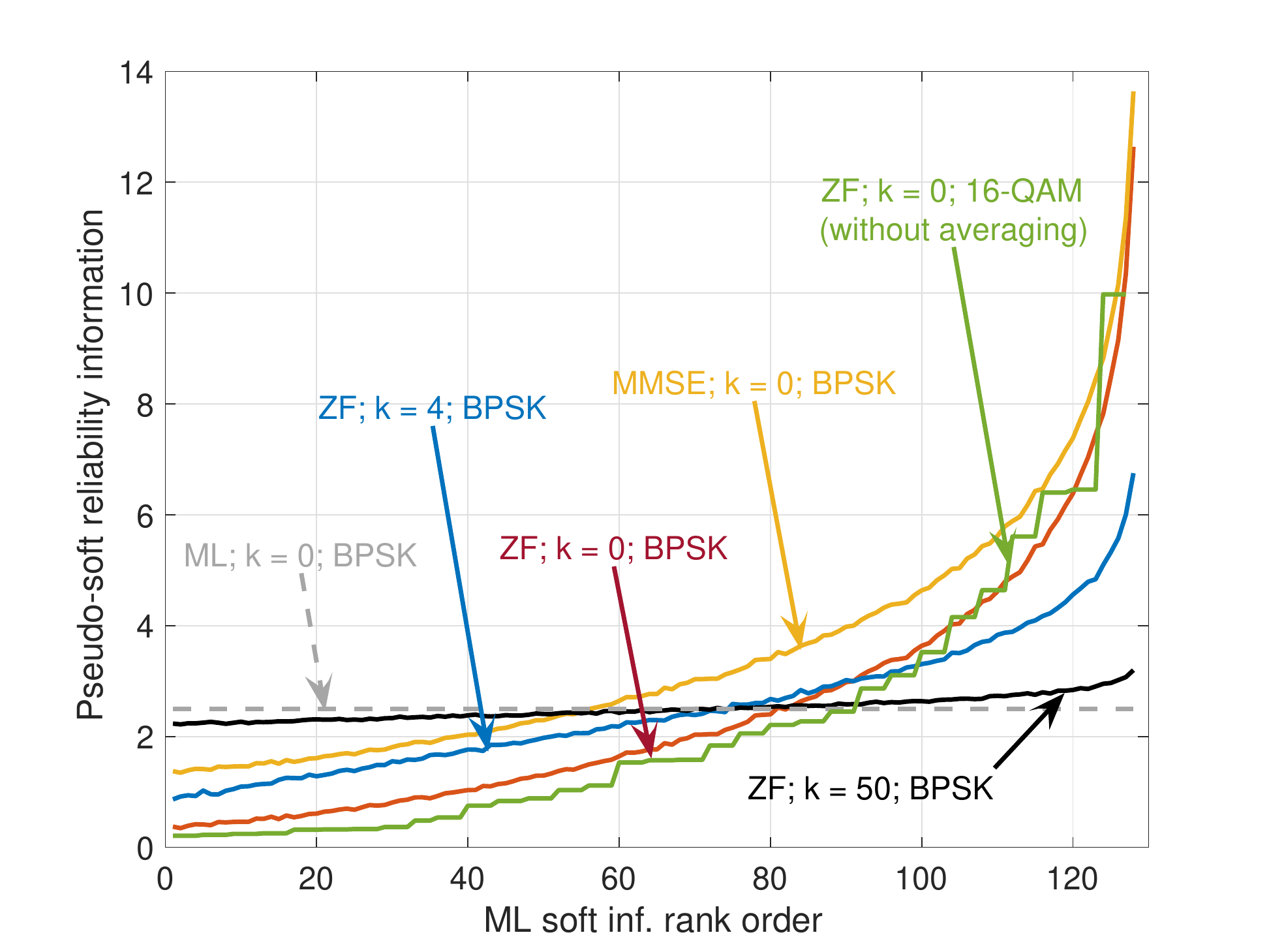}}
  \caption{Distributions of soft and pseudo-soft information under different fading environments.}
  \label{fig:soft_noise}
\end{figure*}

\subsection{Pseudo-soft reliability information in fading channels}
\label{sec:noise_inf}

Although generating soft detection outputs enhances the decoding performance, solely generating hard detection outputs is desirable. The search routines in \eqref{eq:LLR_ML1}, \eqref{eq:LLR_ML2}, \eqref{eq:LLR_ML3}, \eqref{eq:LLR_ZF}, or \eqref{eq:LLR_MMSE} can be computationally demanding, with different complexity costs in each. Furthermore, the $\lambda_{i,j}$ values need to be quantized, say with a five-bit resolution, and passed to the decoder, consuming significantly more bandwidth compared to processing single-bit hard outputs. The hard detectors compute
\begin{equation}\label{eq:hard_det}
\hat{x}_i^{\ML} \!=\! \argmin_{x_i \in \mathcal{X}_i}\abs{y_i - h_i x_i}, \ \
\hat{x_i}^{\MMSE/\ZF} = \left\lfloor \hat{y}^{\MMSE/\ZF}_i \right\rceil_{\mathcal{X}_i},
\end{equation}
where $\lfloor \psi \rceil_{\mathcal{X}_i} \triangleq \argmin_{x_i \in \mathcal{X}_i} \abs{\psi-x_i}$ is a slicing operator that executes simple comparative operations over $\mathcal{X}_i$. The hard-output, $\hat{\mbf{c}}$, is recovered from $\hat{\mbf{x}}$ through demapping.

Alongside $\hat{\mbf{c}}$, we propose generating marginal pseudo-soft information, $\tilde{\mbf{\Lambda}}$, that does not need to be computed on every channel use. In particular, we treat the altered SNR per bit after detection as CAN information. With ML detection, 
\begin{equation}\label{eq:ML_noise}
\tilde{\mbf{\Lambda}}^{\ML} = \frac{1}{\mbf{v}} = \left[\frac{1}{\tilde{\sigma}_{1,1}^2}, \cdots,\frac{1}{\tilde{\sigma}_{i,j}^2}, \cdots, \frac{1}{\tilde{\sigma}_{M,q}^2}\right]\in{\RR^+}^N
\end{equation}
holds information on original CAN only (or original additive white Gaussian noise (AWGN) if, $\forall i$, $\sigma_i\!=\!\sigma$), where $\tilde{\sigma}_{i,j}\!=\!\sigma_{i}$ $\forall j\!\in\!\{1,\cdots,q_i\}$. However, assuming perfect knowledge of the channel coefficients, a channel-induced CAN is recovered from each of the ZF and MMSE detectors, where $\tilde{n}_i^{\ZF}=h_i^{-1}n_i$ and $\tilde{n}_i^{\MMSE}=\left(h_i^{*} h_i\!+\!\sigma_i^{2}\right)^{-1}h_i^*n_i$. The scaled variances, ${\sigma_{i}^{\ZF}}^2$ and ${\sigma_{i}^{\MMSE}}^2$ thus hold both channel and noise information that can approximate soft information. Therefore,
\begin{equation}\label{eq:pseudonoise}
\tilde{\mbf{\Lambda}}^{\ZF/\MMSE} \!=\! \left[\!\frac{1}{\left(\!\tilde{\sigma}_{1,1}^{\ZF/\MMSE}\!\right)\!^2} \!\cdots\!\frac{1}{\left(\!\tilde{\sigma}_{i,j}^{\ZF/\MMSE}\!\right)\!^2} \!\cdots\! \frac{1}{\left(\!\tilde{\sigma}_{M,q}^{\ZF/\MMSE}\!\right)\!^2}\!\right],
\end{equation}
where $\tilde{\sigma}^{\ZF/\MMSE}_{i,j}\!=\!\sigma_{i}^{\ZF/\MMSE}$, $\forall j\!\in\!\{1,\cdots,q_i\}$.

\subsection{Feeding soft or pseudo-soft information to GRAND}
\label{sec:ORBGRAND}
    
Towards understanding the use of soft or pseudo-soft information in GRAND, we express the a posteriori probability that the detected bit $\hat{c}_{i,j}$ is erroneous as 
\begin{align}\label{eq:errprob}
    &\Prob\left( \hat{c}_{i,j} \!\neq\! c_{i,j} | y_i,h_i\right) = \frac{\Prob\left( \hat{c}_{i,j} \!\neq\! c_{i,j}, y_i,h_i \right) }{\Prob\left(y_i,h_i\right)} \\& = \frac{\Prob\left( \hat{c}_{i,j} \!\neq\! c_{i,j} , y_i,h_i \right)/\Prob\left( \hat{c}_{i,j} \!=\! c_{i,j} , y_i,h_i \right)}{1+\Prob\left( \hat{c}_{i,j} \!\neq\! c_{i,j} , y_i,h_i \right)/\Prob\left( \hat{c}_{i,j} \!=\! c_{i,j} , y_i,h_i \right)}.
\end{align}
Noting that $\lambda_{i,j}\!\approx\! \log \left(\Prob\left( c_{i,j}\!=\!1, y_i,h_i \right)\!/\!\Prob\left(c_{i,j}\!=\!0, y_i,h_i \right)\right)$, the probability of bit error can be expressed as
\begin{equation}\label{eq:errprob2}
    \Prob\left( \hat{c}_{i,j} \!\neq\! c_{i,j} | y_i,h_i \right)  = \frac{e^{-\abs{\lambda_{i,j}}}}{1+e^{-\abs{\lambda_{i,j}}}} \in [0,1/2],
\end{equation}
where larger LLR values result in lower bit-flip probabilities. Hence, the a posteriori probability of a noise sequence is approximately \cite{Duffy9414615}
\begin{align}\label{eq:noiseprob}
    \Prob\left( \mbf{w} = \hat{\mbf{w}} \right)  & \approx \prod_{i,j:\hat{w}_{i,j}=0}\left(1-\Prob\left( \hat{c}_{i,j} \!\neq\! c_{i,j} | y_i,h_i \right)\right)\times\\&\prod_{i,j:\hat{w}_{i,j}=1}\Prob\left( \hat{c}_{i,j} \!\neq\! c_{i,j} | y_i,h_i \right) \propto e^{-\sum_{i,j}\abs{\lambda_{i,j}}\hat{w}_{i,j}},
\end{align}
with equality when bits are independent (in the case of interleaving). The likelihood of a putative noise sequence, $\hat{\mbf{w}}$, is thus dictated by the sum of reliabilities of flipped bits, $\eta\left(\hat{\mbf{w}}\right) \!\approx\!\sum_{i,j}|\lambda_{i,j}|\hat{w}_{i,j}$, where less-likely noise sequences have a higher $\eta$. Note that this metric is effectively expressed in Algorithm \ref{alg:SGRAND} using the function $\text{sort}\left(\mbf{W}\times\abs{\mbf{\Lambda}}\right)$.

With knowledge of soft or pseudo-soft information, SGRAND generates $\mbf{w}$ vectors with increasing reliability sums, $\eta\left(\mbf{w}\right)$. The approximation in ORBGRAND, however, depends on the structure of information in $\mbf{\Lambda}$ or $\tilde{\mbf{\Lambda}}$. The quality of information depends on various parameters, such as the noise and channel distributions and the SNR. We investigate the effect of fading on soft information in Fig. \ref{fig:soft_noise}. We consider the Rayleigh channel coefficient, $h^{\text{Ra}}_i$, to be a complex-Gaussian random variable of unit variance $\left(\mathsf{E}[h_i^{{\text{Ra}}^*} h^{\text{Ra}}_i]\!=\!1\right)$, and we define a generalized normalized Rician channel coefficient,
\begin{equation}\label{eq:rician}
    h^{\text{Ri}}_i = \sqrt{k/(1+k)} + \sqrt{1/(1+k)}h^{\text{Ra}}_i.
\end{equation}
On the one hand, a larger $k$ indicates a more deterministic channel component, typically caused by line-of-sight dominance; the channel reduces to AWGN ($h_i \!=\! 1$, $\forall i$) for a very large $k$. On the other hand, a smaller $k$ indicates rich small-scale fading, typically caused by a rich scattering environment; the channel reduces to a Rayleigh fading scenario for $k\!=\!0$.

\begin{figure*}[!ht]
    \centering
    \subfloat[ORBGRAND and CA-SCL decoding of CA-Polar {[\ \!128,105]\ \!} codes under Rayleigh fading, MMSE detection, and BPSK.]{ \label{fig:BLERPolar}\includegraphics[width=0.48\linewidth]{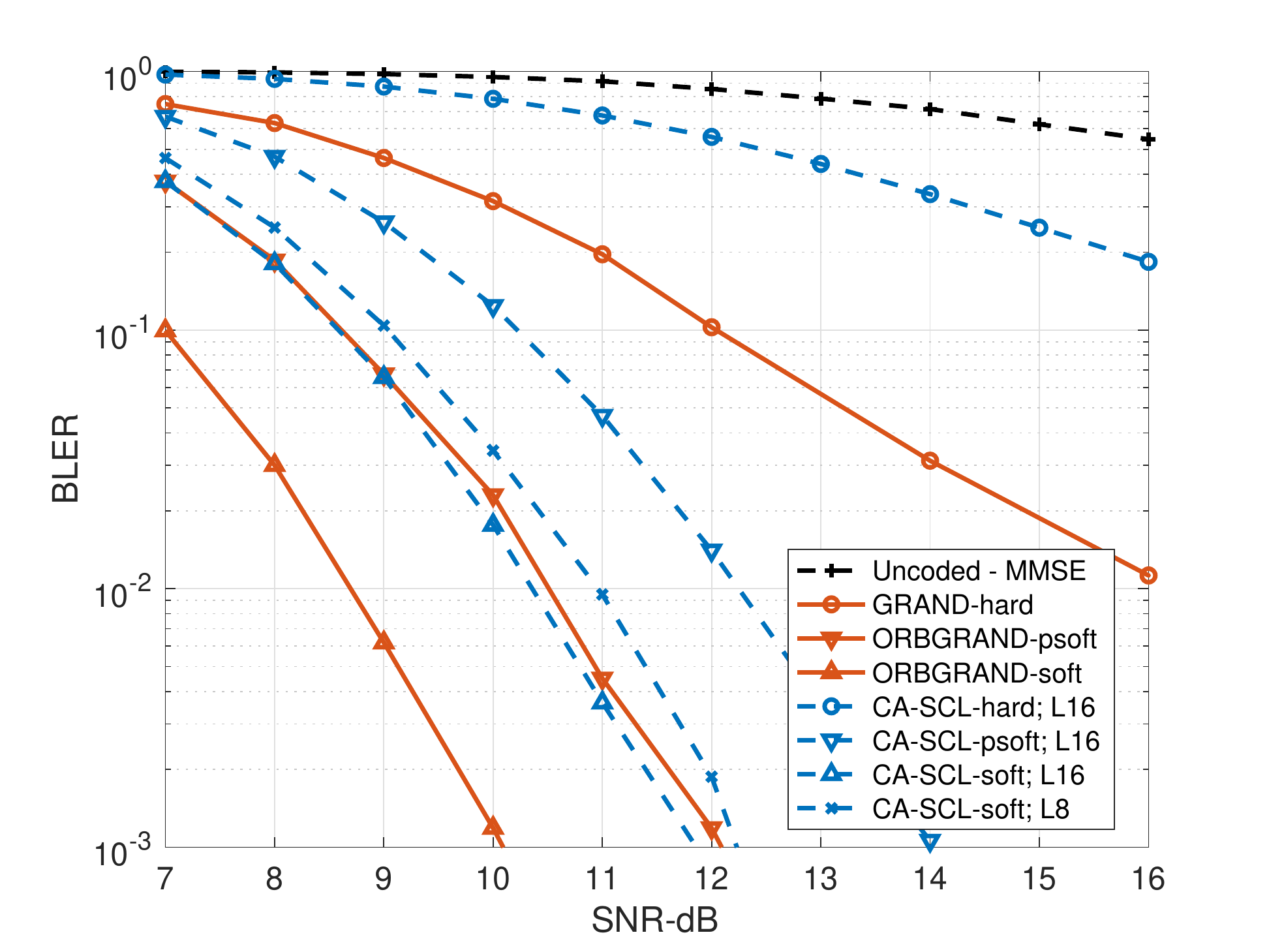}}
    \hfill
    \subfloat[ORBGRAND decoding of BCH {[\ \!113,127]\ \!} codes under Rayleigh and Rician channels, ZF detection, and BPSK. ]{\label{fig:BLERRician}\includegraphics[width=0.48\linewidth]{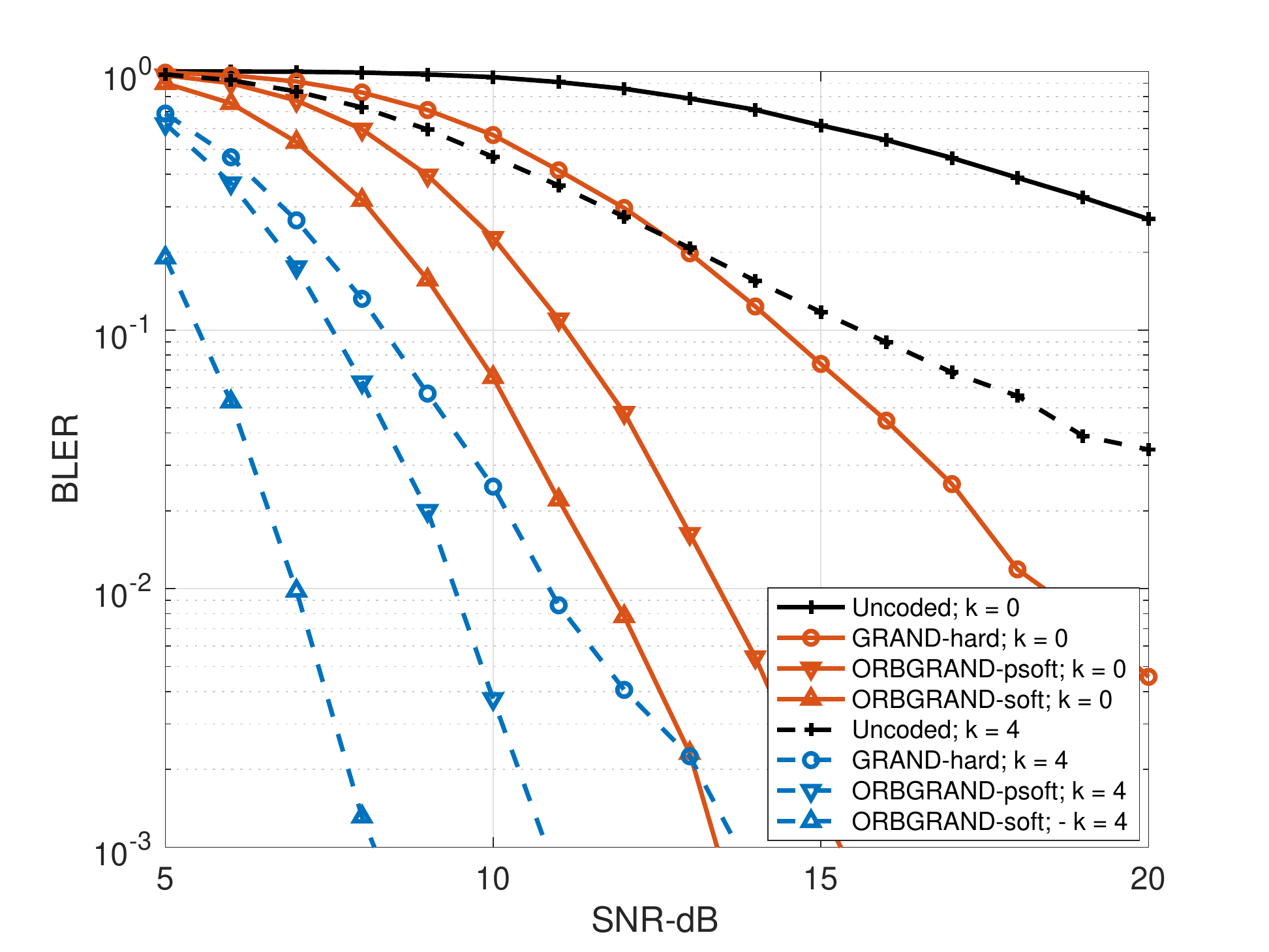}}
  \caption{BLER performance versus SNR under different fading environments.}
  \label{fig:BLER}
\end{figure*}

The soft information of ZF, MMSE, and ML detectors are illustrated in Fig. \ref{fig:llrrician}, for an SNR of $\unit[10]{dB}$. For clarity of presentation, we sort the bit positions, via a reversible permutation, in increasing order of ML reliabilities, where $|\lambda_{i,j}^{\ML}|\leq |\lambda_{\bar{i},\bar{j}}^{\ML}|$ for $i \!\leq\! \bar{i}$, or $i \!=\! \bar{i}$ and $j \!\leq\! \bar{j}$. 
Regardless of the fading type, ML, MMSE, and ZF equalizers generate equally reliable soft-information in point-to-point links. For a large  $k\!=\!50$, an LLR curvature is observed, especially over the least reliable bits, which happen to be the most critical for generating accurate query orders. Such curvature is also noted in \cite{Duffy9414615} for real-valued AWGN channels, where LLRs can be approximated as $\lambda_{i,j}\!=\!y_i$, and $|\lambda_{i,j}|$ follows a folded normal distribution. The higher the SNR, the more noticeable the curvature is, and the higher the soft and pseudo-soft information dynamic range. For our system model, we note from \eqref{eq:LLR_ML3} that $|\lambda_{i,j}|$ is a scaled difference of two chi-squared-distributed random variables, a variance-gamma distribution of slightly slower decreasing tails than the normal distribution. Nevertheless, with rich fading (especially at a low SNR), the reliability curves are almost linear with a zero intercept. %Note that such linearity is more emphasized at lower SNRs. 

The corresponding pseudo-soft information is plotted in Fig. \ref{fig:noiserician}. Contrary to soft information, which is richer with deterministic AWGN channels, pseudo-soft information, dictated by ${\sigma_{i}^{\ZF/\MMSE}}$, is richer under fading. For small $k$ values, ZF and MMSE generate pseudo-soft information that mimics the soft information distribution. However, pseudo-soft information is almost constant with near-deterministic channels. Because ML detection does not apply filtering, its resultant pseudo-soft information is a scalar (nil information) equal to the original SNR before detection. The proposed pseudo-soft information is thus useful with the much less complex linear detectors. Figure \ref{fig:noiserician} further illustrates how the quality of pseudo-soft information degrades with larger modulation types. A stair-step behavior is noted because equalization filters only color noise per symbol; noise information is constant over groups of four consecutive bits with 16-QAM.

Different approximations to the reliability curves have resulted in different ORBGRAND implementations \cite{Duffy9414615}, each providing different decoding complexity and performance tradeoffs. In this work, we adopt the simplest linear approximation in basic ORBGRAND because most of the observed soft and pseudo-soft information with fading are fairly linear, especially over the range of lower-reliability bits. In particular, a linear approximation to $\lambda_{i,j}$ and $\tilde{\lambda}_{i,j}$ is
\begin{equation}\label{eq:appORB}
    \hat{\lambda}_{i,j} \approx \beta \left(iq+j \right).
\end{equation}
The corresponding noise-sequence likelihood is computed as \begin{equation}\label{eq:appORB2}
    \eta\left(\mbf{w}\right) \approx \sum_{i,j:w_{i,j}=1} \hat{\lambda}_{i,j} = \beta \sum_{i,j} \left(iq+j \right) w_{i,j},
\end{equation}
which is proportional to the logistic weight of $\mbf{w}$.

\section{Performance Evaluation}
\label{sec:results}

For simulations, we follow the system model of Sec.~\ref{sec:sysmodel}, considering $\mbf{n}$ to originally be an AWGN \big($\mathsf{E}[\mbf{n}\mbf{n}^{*}]\!=\!\sigma^{2}\mbf{I}_M$\big). Noise coloring is solely introduced by equalization filters at the baseband. Although GRAND works with any block-code construction, we benchmark the achievable gains to state-of-the-art soft CA-SCL decoding of CA-Polar codes and ORBGRAND decoding of Bose–Chaudhuri–Hocquenghem (BCH) codes. We study the block-error rate (BLER) performance over Rayleigh and Rician point-to-point channels. 

We first compare the performance of MMSE-based soft and pseudo-soft (psoft) ORBGRAND schemes to reference CA-SCL soft decoders of CA-Polar {[\ \!128,105]\ \!} codes, alongside basic GRAND and uncoded BLER in Fig.~\ref{fig:BLERPolar}. The proposed pseudo-soft ORBGRAND significantly outperforms hard GRAND (multiple-dB gains) and matches the performance of CA-SCL, list size $L\!=\!16$, with soft information (up to $\unit[0.2]{dB}$); it outperforms soft CA-SCL of list size $L\!=\!8$. Considering the significant reduction in complexity both in ORBGRAND compared to CA-SCL \cite{Duffy9414615} and in processing pseudo-soft compared to soft information, the performance gains are significant. The observed gains are even more significant when considering that foregoing bit interleaving or decreasing the list size further degrades CA-SCL decoding. In contrast, the performance of GRAND-based decoders can be enhanced by leveraging noise coloring in the absence of interleavers and whitening filters. Similar gains are demonstrated in Fig. \ref{fig:BLERRician} for BCH {[\ \!113,127]\ \!} codes, where ORBGRAND with ZF-induced pseudo-soft information outperforms hard GRAND by around $\unit[7]{dB}$ at a BLER of $10^{-3}$, while only lagging by $\unit[2]{dB}$ behind soft ORBGRAND, which is a benchmark soft decoder of BCH codes. Therefore, fading-induced statistical noise information is beneficial for efficient GRAND realizations.

\begin{figure}[!t]
  \centering
  \includegraphics[width=\linewidth]{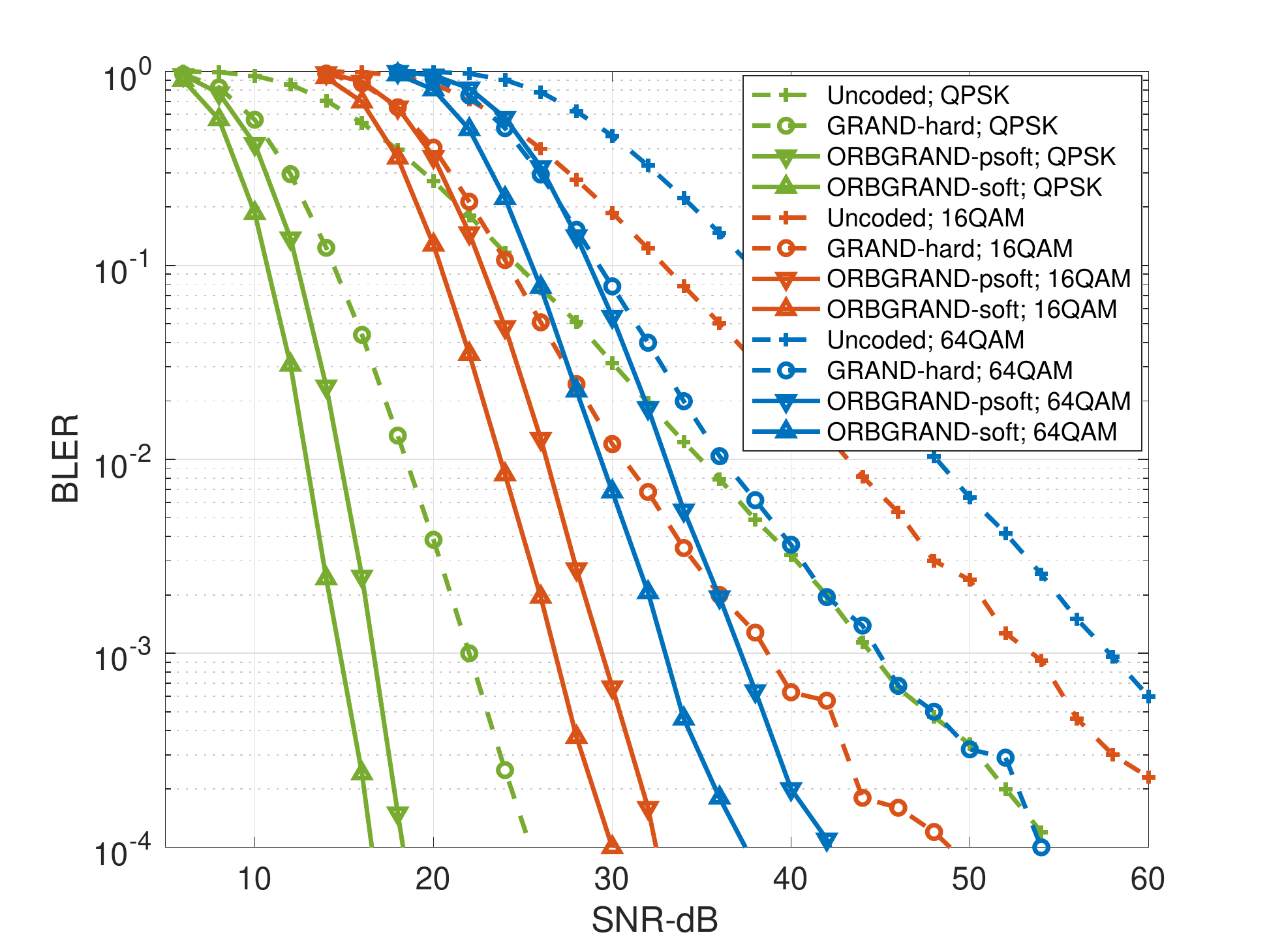}
  \caption{BLER of ORBGRAND decoding of BCH {[\ \!113,127]\ \!} codes under Rayleigh fading, ZF detection, and higher-order QAMs.}\label{fig:BLER-modulations}
\end{figure}

The performance under Rician fading ($k\!=\!4$) is also illustrated in Fig. \ref{fig:BLERRician}. As argued in Sec.\ref{sec:ORBGRAND} (Fig.\ref{fig:soft_noise}), with larger $k$ values, the point-to-point channel is better-conditioned and soft information improves; even pseudo-soft ORBGRAND improves because the hard demapped symbol vector would have fewer errors. However, compared to soft information under same fading conditions, the gap in pseudo-soft information increases with $k$; a $\unit[3]{dB}$ gap at $k\!=\!4$ compared to a $\unit[2]{dB}$ gap at $k\!=\!0$ (Rayleigh fading). Therefore, as expected, diminishing gains of pseudo-soft information are noted with Rician fading. 

By increasing the modulation order to quadrature phase-shift keying (QPSK), 16-QAM, and 64-QAM, Fig. \ref{fig:BLER-modulations} illustrates that the gain in pseudo-soft information compared to hard decoding increases (more than $\unit[10]{dB}$ at a BLER of $10^{-3}$). However, compared to the ORBGRAND that avails of full soft information, the gap also increases (up to $\unit[4]{dB}$). The latter is caused by the increased correlation in pseudo-soft information (the stepped pattern in Fig. \ref{fig:noiserician}). The coarse quantization of pseudo-soft information with higher-order modulations is particularly harmful to basic ORBGRAND with linear approximation. To remedy this effect, instead of arbitrary breaking ties over the $q$ $\tilde{\lambda}_{i,j}$ values for a specific $i$, we can consider the absence of soft information at the symbol-bit level and adopt an increasing Hamming weight metric (combination of logistic and Hamming weight metrics). In general, for a probability of bit flip less than $1/2$, in the absence of soft information or further channel knowledge, noise query follows increasing Hamming weights, as in the case of hard-detection GRAND. Adopting the quantized GRAND algorithm that avails of any level of quantized soft information \cite{duffy2022quantized} is also promising with coarse pseudo-soft information. 

%Note that the pattern in Fig. \ref{fig:BLER-modulations} will emerge after sorting bits, regardless of the presence of interleavers, further highlighting the redundant use of interleavers with GRAND.

%\red{Figure \ref{fig:ORBvsSGRAND} further compares the ORBGRAND performance under Rician ($k\!=\!5$) channels to that of optimal SGRAND. With noise-information, SGRAND has a small advantage over ORBGRAND, suggesting that the linear approximation of basic ORBGRAND does not properly capture the curvature of noise-information at high-reliability bits (Fig. \ref{fig:noiserician}). }

%\red{The lack of curvature at lower-reliability bits in noise-information plots explains the increased gap compared to soft-information BLER under limited guess budgets. Also, the larger the $K$ factor (the closer to AWGN), the more flat noise-information gets (the less useful).}

\section{Conclusions}
\label{sec:conc}

We proposed a framework for soft GRAND decoding under fading channel conditions. By leveraging low-cost channel state information and statistical noise information in pseudo-soft bit reliability metrics, pseudo-soft GRAND schemes are shown to contend with benchmark state-of-the-art decoding schemes that avail of higher-cost complete soft information, introducing up to $\unit[10]{dB}$ SNR gains at a $10^{-3}$ BLER over hard-GRAND. This work can extend into a generic joint detection and decoding framework in which equalization filters are designed to intelligently avail of the structure of noise in order to optimize data detection and GRAND guesswork jointly.


\end{document}